\documentclass{PoS}

\usepackage{aas_macros}
\usepackage{tabu}
\usepackage{overpic}
\usepackage{multirow}
\usepackage{paralist}
\usepackage{rviewport}
\usepackage{wrapfig}
\usepackage{multicol}
\usepackage[font=footnotesize,labelfont=bf]{caption}

\title{Uncertainties in the Magnetic Field of the Milky Way}

\ShortTitle{Uncertainties in the Magnetic Field of the Milky Way}

\author{\speaker{Michael Unger}\\ Institute for Nuclear Physics,
  Karlsruhe Institute of Technology, 76344 Eggenstein-Leopoldshafen,
  Germany\\ E-mail: \email{michael.unger@kit.edu}}
\author{Glennys R.\ Farrar\\ Center for Cosmology and Particle
  Physics, New York University, New York, NY 10003, USA\\ E-mail:
  \email{gf25@nyu.edu}}

\abstract{We improve on the model of the Galactic Magnetic Field (GMF) from Jansson \& Farrar (2012), which was constrained using all-sky rotation
measures of extragalactic sources and polarized and unpolarized synchrotron emission data from WMAP.   We have
  \begin{itemize}
     \item developed several alternative functional forms for
            the coherent and random components
     \item used newer synchrotron products from Planck and WMAP
     \item used new models of the densities of thermal electrons
           and cosmic-ray electrons.
  \end{itemize}
The differences in the resultant GMF models, depending on which
parameterization of the field, synchrotron product and electron
densities are used, provides a measure of the uncertainty in our
inference of the GMF.  We discuss the impact of these uncertainties on
charged-particle astronomy at ultra-high energies.}

\FullConference{35th International Cosmic Ray Conference --- ICRC2017\\
		10--20 July, 2017\\
		Bexco, Busan, Korea}

\begin{document}

\section{Introduction}

Spiral galaxies such as the Milky Way are known to be permeated by
magnetic fields (see e.g.~\cite{2004astro.ph.11739S,
  2015ASSL..407..483H, 2016A&ARv..24....4B}).  Global models of these
fields are frequently used to study a variety of astrophysical
phenomena including star formation, gas dynamics, the propagation and
acceleration of low-energy cosmic rays
and the deflections of the arrival directions of ultrahigh-energy cosmic rays.

The hitherto most complete attempt to determine the global structure
of the Galactic magnetic field (GMF) is the model of
Jansson \& Farrar~\cite{2012ApJ...757...14J, 2012ApJ...761L..11J}
(JF12). In this model, the GMF is described by a superposition of
three divergence-free  large-scale regular components: a spiral disk field, a
toroidal halo field and a poloidal field (``X-field''). In
addition, there is a turbulent field model whose disk  component is modeled following
the same spiral structure as the regular component and there is also
an extended random halo field. The 22+14 parameters of the regular
and random magnetic field are constrained by: \vspace*{2mm}
\begin{compactitem}
  \item[(a)] Multi-frequency radio observations of the Faraday
    rotation of extragalactic radio sources. The corresponding
    {\itshape rotation measures} (RMs) are proportional to the
    line-of-sight integral of the longitudinal magnetic field,
    weighted with the density of thermal electrons of the warm ionized
    medium of the Galaxy.
  \item[(b)] Measurements of the polarized synchrotron emission of
    cosmic-ray electrons in the regular magnetic field of the
    Galaxy. The total polarized intensity (PI) is proportional to the
    line of sight integral of the ordered component of the transverse
    magnetic field strength, weighted by the density of cosmic-ray
    electrons. The direction of the transverse magnetic field
    component can be inferred from the Stokes parameters Q and U.
  \item[(c)] Measurements of the total (polarized and unpolarized)
    synchrotron intensity I, which is a line-of-sight integral
    depending on the product of cosmic-ray electron density and total
    transverse magnetic field strength (coherent and random).
\end{compactitem}\vspace*{2mm}

In this contribution we investigate the uncertainties in our inference
of the GMF introduced by the various assumptions entailed in the
global modeling of the field, especially the functional forms used for
the field characterizations and the models for the thermal and
cosmic-ray electrons. In addition, we consider uncertainties due to
uncertainties in the observables used to constrain the field model.

After describing the data analysis procedure in the next section,
various model variations will be discussed in Sec.~\ref{sec:mv}. The
finite number of variations investigated can of course not provide an
exhaustive investigation of models compatible with the data, therefore
the results presented in this article provide a {\itshape lower limit}
on the uncertainties of the GMF, in the absence of further input to
select among or discard some of the GMF model variations.  As an
example of the impact of the uncertainties, we discuss in
Sec.~\ref{sec:cpa} the uncertainties of the arrival directions of
ultrahigh energy cosmic rays due to the deflections in an uncertain
Galactic magnetic field.

The analysis presented here follows closely the procedure developed
for the JF12 model. We use the same catalogue of 40403 extragalactic
RMs and bin it into 3072 equal-area HEALPix pixels on the sky to
define the mean and variance of the RMs in a particular direction. A
similar procedure is applied to the synchrotron maps from WMAP and
Planck. For each model of the GMF we produce predictions of the
observables RM, Q, U and I by a line-of-sight integration of the
magnetic field weighted with the thermal electrons density and
cosmic-ray electrons spectrum respectively.  For this purpose we
developed a new computational framework that allows for adaptive step
control during the integration and a highly modular model
configuration that is needed for the variations described in the next
section. A comparison of the measured observables with the simulated
predictions from the JF12 model is shown in the figure
below.

\begin{figure}[h!]
\centering
\includegraphics[width=0.9\linewidth]{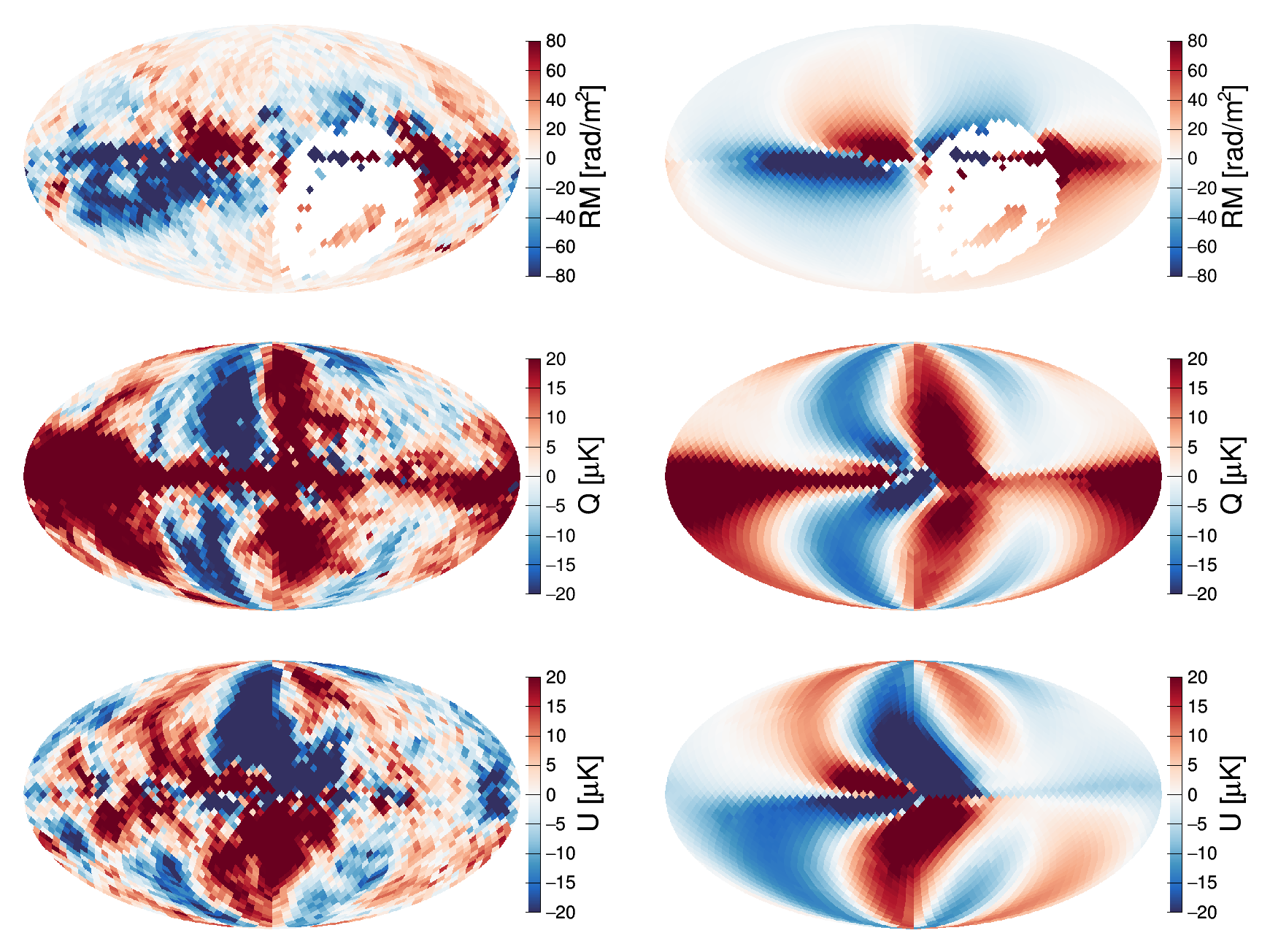}\\
\includegraphics[width=0.9\linewidth]{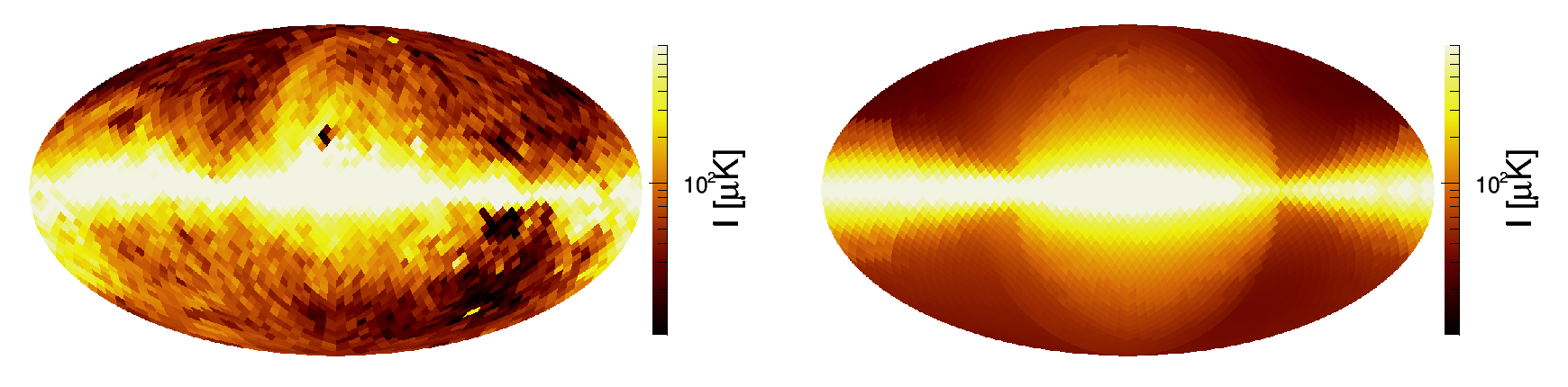}
\caption{Measured (left column) and simulated (right column) sky maps
  of rotation measures (top), polarized synchrotron emission (Stokes Q
  and U parameters in the two middle panels) and total synchrotron
  intensity (bottom). The synchrotron data are
  from~\cite{2011ApJS..192...15G} and the GMF model is
  JF12~\cite{2012ApJ...757...14J,2012ApJ...761L..11J}.}
\label{fig:jf12}
\end{figure}

\section{Data Analysis}

The parameters of the different GMF models are optimized by minimizing
the variance-weighted pixel-by-pixel difference between the predicted
and measured observables. In~\cite{2012ApJ...757...14J}, this
optimization was performed with a time-intensive MCMC sampling of the
posterior distribution of the model parameters. The model variations
investigated in the present work usually lead to much larger differences in
the estimates of the GMF than the single-model statistical
uncertainties. Therefore the posterior distribution of the parameters
are of secondary importance and the $n$ model parameters can be
optimized more efficiently with multi-dimensional gradient
search~\cite{1975CoPhC..10..343J}.  For each model variation, the
gradient search is perform typically 50 times in parallel starting at
random positions in the $n$-dimensional parameter space to assure that
the global minimum is found.


\section{Model Variations}
\label{sec:mv}
In the following we will give a brief overview of the investigated
modeling variations.  The different models are listed in
Table~\ref{tab:cohmodels}. As can be seen in the last column of this
table, all of the 20 modeling variations give a reasonable description of
the data with reduced $\chi^2$s ranging from 0.78 to 1.22. Note that
for all models we use the same relative weighting of the $\chi^2$
contributions of the RM and synchrotron data. Moreover, the same pixel
mask is applied in all cases (see~\cite{2012ApJ...757...14J} for a
description of the mask).  It should be emphasized that due to differences in the data processing (especially in the synchrotron data products), the reduced $\chi^2$ is not always appropriate for comparing the fit quality in the model variations, as discussed below.

\paragraph{Parameterizations}
We tested several alternatives to the parameterizations of the coherent
field used in the JF12 model. All these alternatives are
divergence-free, like the original functional forms. First, we replaced
the parametric X-field of JF12 with ``model C''
from~\cite{2014A&A...561A.100F} (variation $b$). The main difference
between these two models is that in the former the angle of the
poloidal field lines with respect to the Galactic plane is kept
constant above a certain galacto-centric radius, whereas in ``model
C'' this angle evolves with radius and height.  (The smoother behavior of ``model C" at the Galactic plane is unobservable
due to random fields.)  In the next step we
restrict the radial extent of the toroidal field to be the same in the
Northern and Southern hemisphere of the Galaxy leading to a best-fit
value of $r_{\rm N/S}=11.3\pm 0.3$~kpc (variation $c$). Introducing a
Galactic warp~\cite{2006ApJ...643..881L} into the model (variation
$d$) leaves most of the line-of-sights unchanged and introduces only
minor changes to the simulated observables at low latitudes close to
the Galactic anti-center. The disk field of JF12 is implemented as a
logarithmic spiral with a pitch angle of 11.5$^\circ$ and eight
discontinuous magnetic field ``arms''
following~\cite{2007ApJ...663..258B}. We replaced this ansatz by a
divergence-free smooth spiral field with free pitch angle (variation
$e$). Interestingly, the best fit gives a pitch angle of $(14.5 \pm
0.6)^\circ$, i.e.\ close to the pitch angles determined for the spiral
segments of the Milky Way (see e.g.~\cite{2014ApJ...783..130R}) in
accordance to observations of external
galaxies~\cite{2016ApJ...833...43C}. Finally, we replaced the independent
toroidal and poloidal components of JF12 by a ``twisted X-field'',
i.e.\ a originally poloidal field that is evolved in time taking into
account the radial and vertical shear of the Galaxy.  Allowing for
different parametric forms of the radial dependence of the
initial field yields model variations $f$ and
$g$~\cite{UF17}.

Despite the considerable differences between these parameterization,
the quality of the fits is similar for each of them
(cf.~Table~\ref{tab:cohmodels}), i.e.\ 
the data does not clearly discriminate between these model variations at this stage.

{
\begin{table}[t]
\small
\extrarowsep=-0.5mm
\centering
\begin{tabu}{c|cccccccc}
  \multirow{2}{*}{id} & disk & toroidal & poloidal & thermal & cosmic-ray & synchrotron & \multirow{2}{*}{misc.}  &\multirow{2}{*}{$\chi^2/$ndf}\\
  & model & model & model & electrons & electrons & data product & &\\\hline
\multicolumn{5}{l}{\vspace*{-.2cm}}\\
\multicolumn{5}{l}{\bf Parametric models}\\
 a & JF & {JF} & {JF} & NE2001 & GP$_{\rm JF}$ & WMAP7 & - & 1.10\phantom{$^*$}\\ 
 b & JF & {JF} & FTC & NE2001 & GP$_{\rm JF}$ & WMAP7 & - & 1.09\phantom{$^*$}\\ 
 c & JF & JFsym & FTC & NE2001 & GP$_{\rm JF}$ & WMAP7 & - & 1.11\phantom{$^*$}\\ 
 d & JF & JFsym & FTC & NE2001 & GP$_{\rm JF}$ & WMAP7 &   warp & 1.11\phantom{$^*$}\\ 
 e &  UF & JFsym & FTC & NE2001 & GP$_{\rm JF}$ & WMAP7 & - & 1.09\phantom{$^*$}\\ 
 f &  UF & \multicolumn{2}{c}{  UFa}  & NE2001 & GP$_{\rm JF}$ & WMAP7 & - & 1.14\phantom{$^*$}\\ 
 g &  UF & \multicolumn{2}{c}{  UFb} & NE2001 & GP$_{\rm JF}$ & WMAP7 & - & 1.09\phantom{$^*$}\\ 
\multicolumn{5}{l}{\vspace*{-.2cm}}\\
\multicolumn{5}{l}{\bf Synchrotron products}\\
 h & JF & JFsym & FTC & NE2001 & GP$_{\rm JF}$ &  WMAP9base & - & 1.22$^\dagger$\\ 
 i & JF & JFsym & FTC & NE2001 & GP$_{\rm JF}$ &  WMAP9sdc & - & 1.24$^\dagger$\\ 
 j & JF & JFsym & FTC & NE2001 & GP$_{\rm JF}$ &  WMAP9fs & - & 1.11$^\dagger$\\ 
 k & JF & JFsym & FTC & NE2001 & GP$_{\rm JF}$ &  WMAP9fss & - & 1.22$^\dagger$\\ 
 l & JF & JFsym & FTC & NE2001 & GP$_{\rm JF}$ &  Planck15  & - & 0.78$^\dagger$\\ 
\multicolumn{5}{l}{\vspace*{-.2cm}}\\
\multicolumn{5}{l}{\bf Thermal electrons}\\
 m & JF & JFsym & FTC &  YMW17 & GP$_{\rm JF}$ & WMAP7 & - & 1.21\phantom{$^*$}\\ 
 n & { UF} & JFsym & FTC &  YMW17 & GP$_{\rm JF}$ &  WMAP7 & - & 1.14\phantom{$^*$}\\ 
 o & JF & {{JF}} & FTC & NE2001 & GP$_{\rm JF}$ & WMAP7 & $\kappa=-1$ & 1.05$^*$\\ 
 p & JF & {{JF}} & FTC & NE2001 & GP$_{\rm JF}$ & WMAP7 & $\kappa=+1$ & 1.05$^*$\\\ 
 q & JF & JFsym & FTC & NE2001 & GP$_{\rm JF}$ & WMAP7 &  HIM & 1.12\phantom{$^*$}\\ 
\multicolumn{5}{l}{\vspace*{-.2cm}}\\
\multicolumn{5}{l}{\bf Cosmic-ray electrons}\\
 r & JF & JFsym & FTC & NE2001 &  O13a & WMAP7 & - & 1.13\phantom{$^*$}\\ 
 s & JF & JFsym & FTC & NE2001 &  O13b & WMAP7 & - & 1.12\phantom{$^*$}\\ 
 t & JF & JFsym & FTC & NE2001 &  S10 & WMAP7 & - & 1.13\phantom{$^*$}\\ 
\end{tabu}
\caption{Summary of model variations investigated in this paper. The
  original JF12 model corresponds to the first row (model a) and the
  reference model is given in the third row (model 3). The goodness of
  fit for describing the RM, Q and U data is given in the last column
  with the exception for the combined fits of coherent and random
  field (marked with a $^*$), where the $\chi^2$ also includes the
  contribution from the total intensity I. The  $\chi^2$s of the fits
  with different synchrotron data products (marked with a $^\dagger$) used
  different weights in the fits derived from these products.}
\label{tab:cohmodels}
\end{table}
}

\paragraph{Synchrotron Data Products}
\label{sec:syn}
As the baseline for the comparisons considered in this work, we use
the 7-year WMAP synchrotron maps ~\cite{2011ApJS..192...15G}, as
originally used in JF12.  But for variations $h$-$l$ we replace that
synchrotron data with different synchrotron products from the 9-year
final WMAP data release~\cite{2013ApJS..208...20B} and the Planck 2015
data release~\cite{2016A&A...594A..10P}. These products differ in the
constraints applied to the measured Galactic microwave emission data
to extract the synchrotron component. WMAP7 and WMAP9base fit the data
with a sum of synchrotron, free-free and dust emission. All other
models include a spinning dust component to describe the ``anomalous
microwave emission''. Moreover, different constraints to the spectral
index of the synchrotron emission are applied in different products.
It must be emphasized that the sizable
differences in the reduced $\chi^2$s for variations $h-l$ visible in
Table~\ref{tab:cohmodels} originate from the different variances
inferred from the sub-pixel variation in the data products, reflecting the smoothing procedure of~\cite{2016A&A...594A..10P}.
Thus the relative reduced $\chi^2$s here are \emph{not} indicative of a better or worse description of the
data.

The differences between the various synchrotron products are small
for the polarized emission and therefore the inferred coherent
magnetic field does not change much when switching from one data
product to another.  However, as already noted
in~\cite{2016A&A...596A.103P}, the different treatment of the
anomalous microwave emission in the models used for the  different synchrotron products strongly affects the estimated total
synchrotron intensities, and thus the random magnetic field component.  This difference
leads to a reduction of random field strength, by up to a factor of
four in the disk, relative to JF12~\cite{UF17}.

\paragraph{Thermal Electrons}

A model of the density of thermal electrons in the Galaxy ($n_e$) is
needed to predict the rotation measures for a given magnetic field
configuration. Estimates of the spatial distribution of $n_e$ rely
mostly on measurements of the dispersion measures of Galactic pulsars
and to a lesser extent on scattering measures of Galactic and
extragalactic sources. We tested the impact of two different models for the thermal
electron densities: NE2001~\cite{2002astro.ph..7156C}, with the updated
scale height of the thick disk from~\cite{2008PASA...25..184G}, and
YMW17~\cite{2017ApJ...835...29Y}. While the newer YMW17 model relies
on more dispersion measures from pulsars with measured distances, the
more important difference between the two models lies in their particular
parametric choices for the model components, such as the thickness
and pitch angles of the spiral arms. Both models give a similar
performance when comparing the predicted to measured
dispersion measures. The slightly worse $\chi^2$ values of the YMW17
variations $m$ and $n$, with respect to the corresponding NE2001 fits $c$ and $e$,
can be partially attributed to worse description of the RMs in the
direction of the Gum nebula when using YMW17. The main difference in
the inferred magnetic fields is a larger field strength in the
halo:  due to the lower density of electrons at large Galactic height
in the YMW17 model, the fitted field strengths for the toroidal and
poloidal components are about a factor two larger than if
determined using NE2001.
We also tested the effect of adding the large-scale hot ionized medium
from~\cite{2015ApJ...800...14M} to the calculation of RMs, but no big effect
on the inferred GMF was found (variation $q$).

In a further $n_e$-related variation, we tested the impact of assuming
a correlation, $\kappa$, between the magnetic field strength and the
thermal electron densities. As argued in~\cite{2003A&A...411...99B},
pressure balance in the magneto-ionic medium can lead to an
anti-correlation of $n_e$ and the magnetic field, leading to an
underestimate of the coherent magnetic field strength inferred from
rotation measures if the anti-correlation is ignored. On the other
hand, compression of the magneto-ionic medium causes an enhancement of
both the magnetic field strength and gas densities, leading to
positive correlation between $n_e$ and the magnetic field resulting in
an overestimate of $B_{\rm coh}$ inferred from RMs with $\kappa = 0$.
We studied the two extreme cases $\kappa=-1$ (variation $o$) and
$\kappa=+1$ (variation $p$) by performing a combined fit of the
coherent and random fields using the modified relations between RM and
the magnetic field given in~\cite{2003A&A...411...99B}.  As expected,
very different coherent magnetic field strengths were found
for the two cases.  The total energy of the coherent field in the
Galaxy is $3.5\times 10^{55}$~erg for $\kappa=-1$ and $3.8\times
10^{54}$~erg for $\kappa=+1$. Under the standard assumption of no
correlation (variation $b$), the coherent energy is $1.2\times
10^{55}$~erg. Note that these are upper bounds on the effects of a
possible $n_e$-$B$ correlation, because a) the assumed correlation
coefficients are at the extreme values and b) the synchrotron product
used in the comparison is the baseline one from WMAP7 without a spinning dust component, whose inferred random magnetic field is largest (cf. previous section).

\paragraph{Cosmic-Ray Electrons}
The density and spectrum of cosmic-ray electrons depends in two ways
on the Galactic magnetic field: Firstly, the GMF determines the
diffusion of the electrons from their sources through the Galaxy and,
secondly, synchrotron losses at high electron energies is the main cause of electron cooling.  The JF12 fits used $n_{\rm cre}$ from a
two-dimensional {\scshape GalProp} simulation with a uniform isotropic
diffusion coefficient within a cylindrical volume of 4~kpc height.
Here we tested the three variants of the cosmic-ray electron densities
used in~\cite{2016A&A...596A.103P}, which are updated versions of the
calculations described in~\cite{2010ApJ...722L..58S}
and~\cite{2013MNRAS.436.2127O}. Two of them (variation $r$ and $t$)
use a vertical extent of the diffusion volume of 4~kpc, whereas for
variation $s$ the height of the diffusion volume is 10~kpc. The
inferred field strengths of the coherent GMF are remarkably robust
under these changes, partially being due to the flexibility in the fit
to adjust the relative scale between the RMs and the polarized
intensity by changing the amount of striated fields, i.e.\ aligned
anisotropic random fields that contribute to the polarized intensity,
but not to the rotation measures. Further studies concerning the
impact of cosmic-ray electrons will be shown in~\cite{UF17}, in
particular the effect of using a more detailed three-dimensional source distribution of
relativistic electrons.

\begin{figure}[t!]
\begin{overpic}[width=0.5\linewidth]{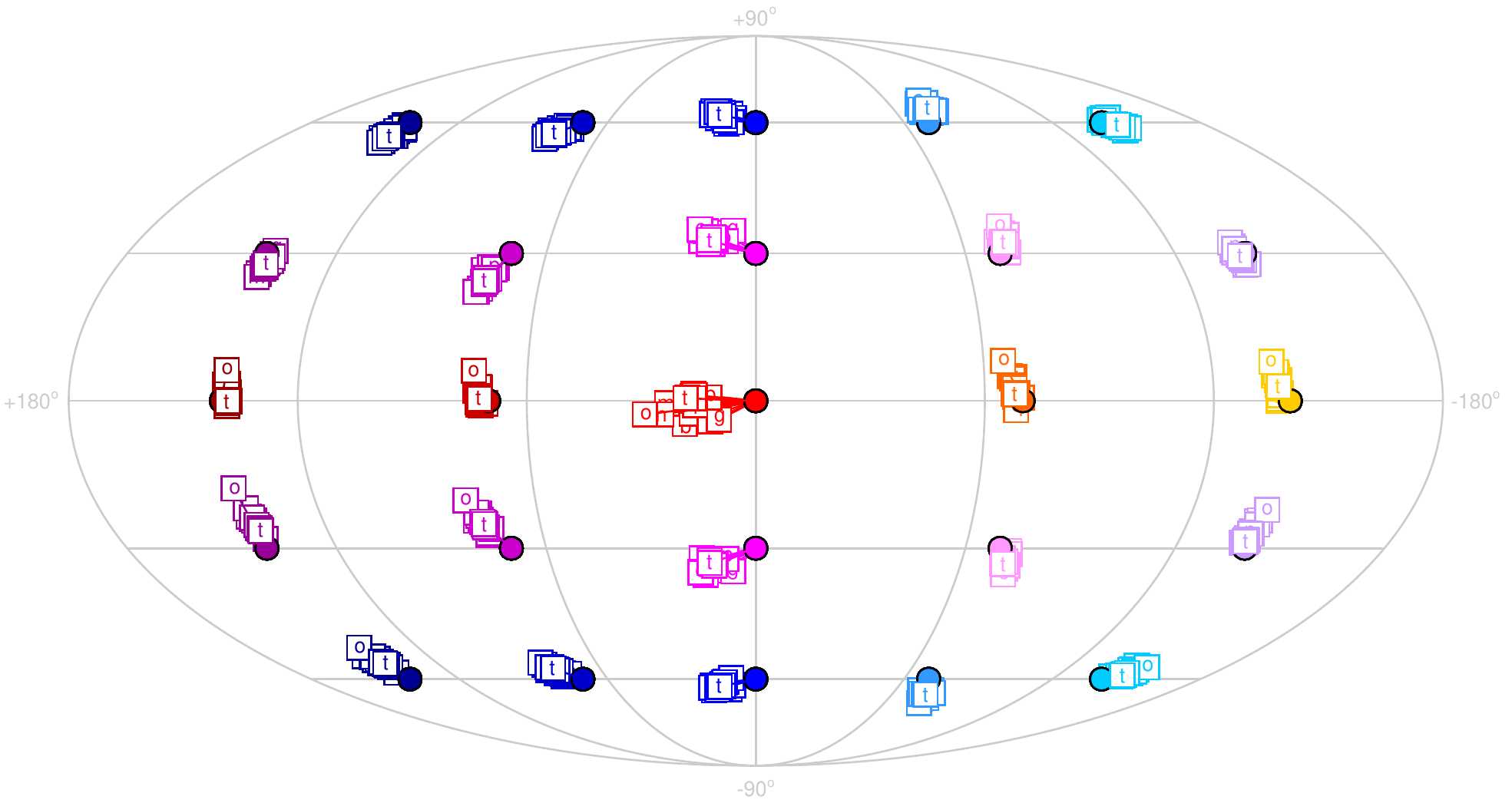}
\put (2,48) {\scriptsize $R = 60$~EV}
\end{overpic}
\begin{overpic}[width=0.5\linewidth]{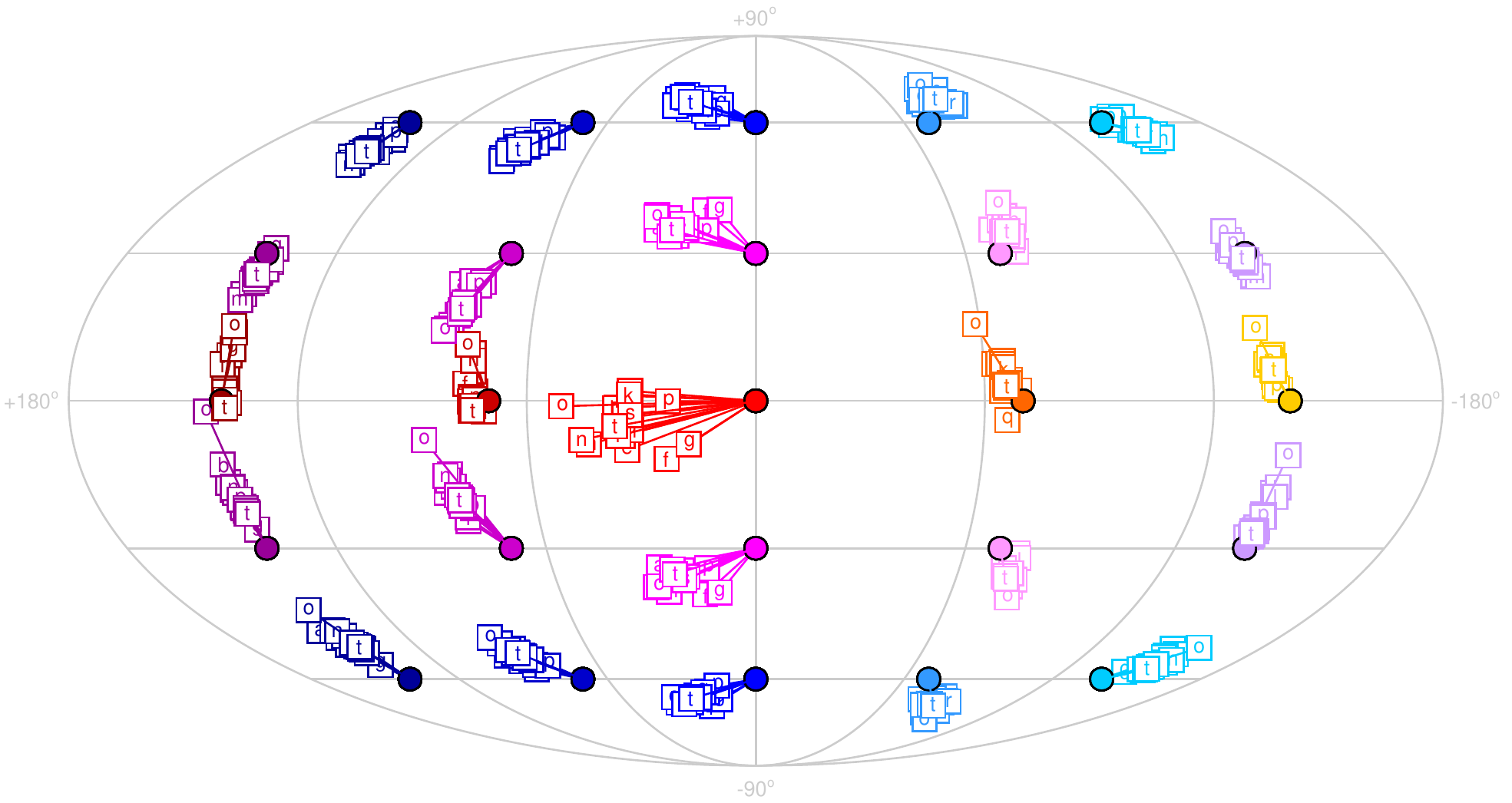}
\put (80,48) {\scriptsize $R = 30$~EV}
\end{overpic}\\
\begin{overpic}[width=0.5\linewidth]{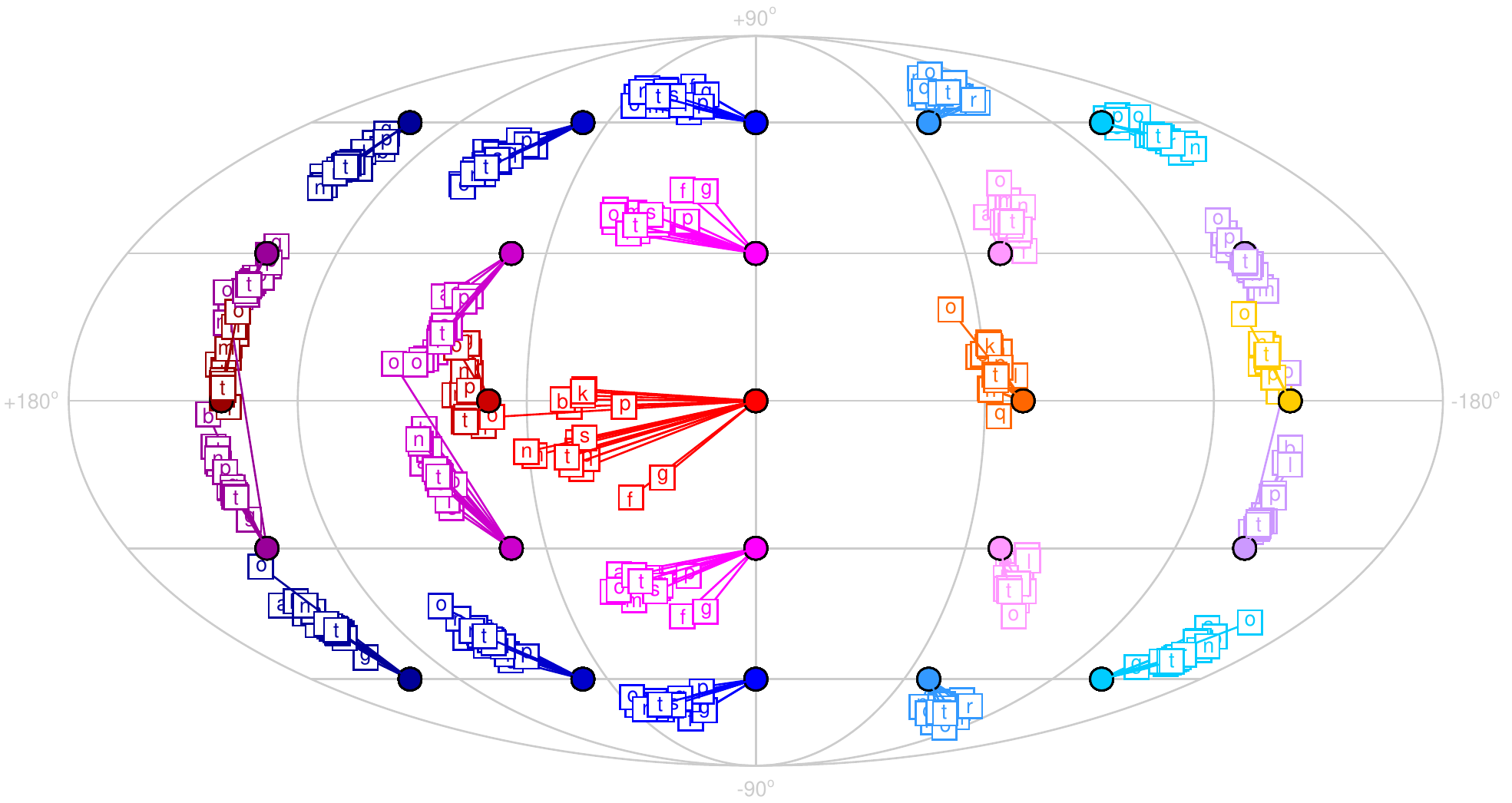}
\put (2,2) {\scriptsize $R = 20$~EV}
\end{overpic}
\begin{overpic}[width=0.5\linewidth]{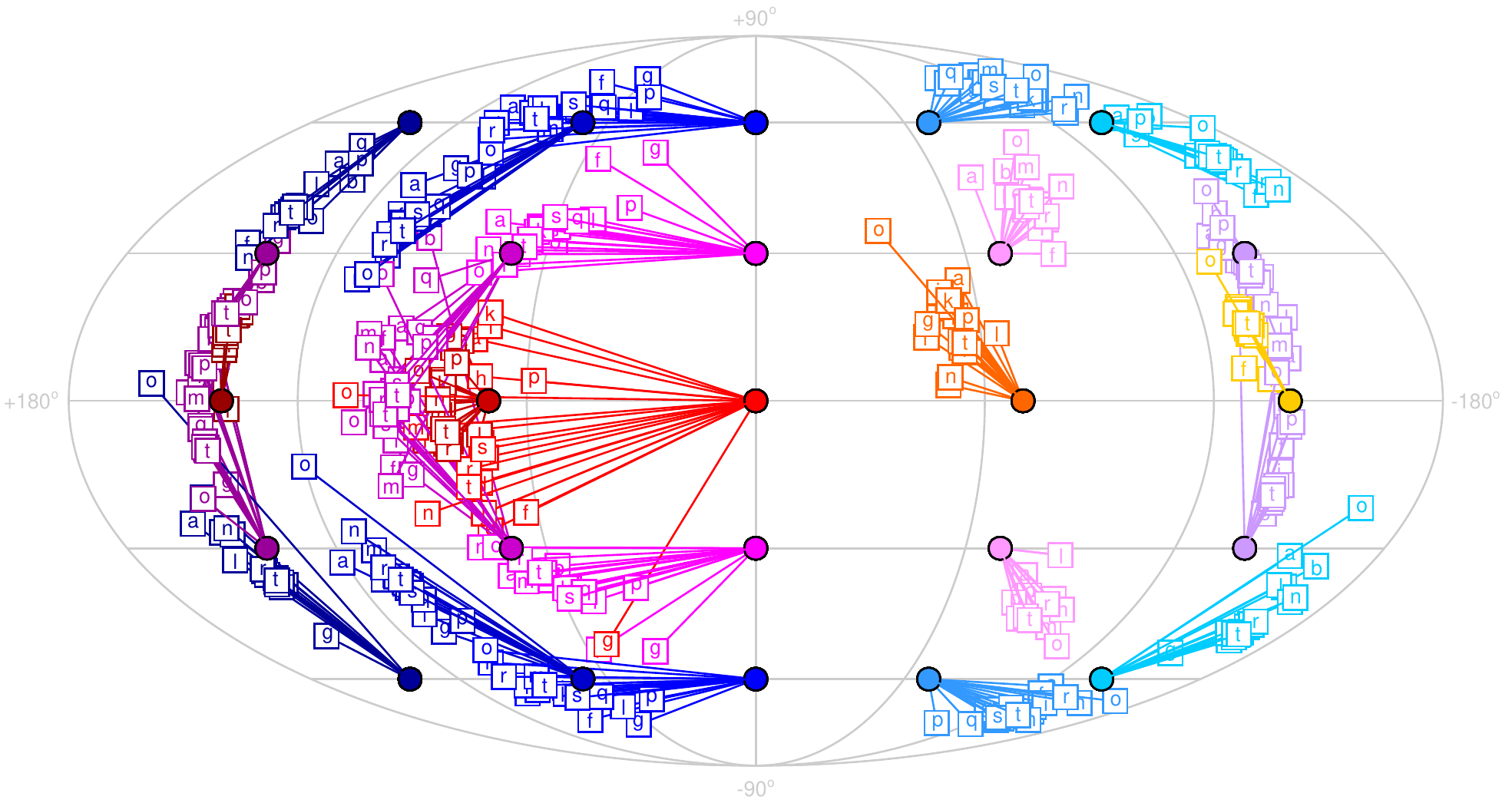}
\put (80,2) {\scriptsize $R = 10$~EV}
\end{overpic}
\caption[backtracking]{Backtracking of charged particles at different
  rigidities from a regular grid of initial directions (dots) through
  different models of the coherent GMF. The resulting directions
  outside of the Galaxy are denoted by squares and the letters
  correspond to the models listed in Table~\ref{tab:cohmodels}. The sky
  maps are in Galactic coordinates and the particle rigidities
  indicated in corners of each panel.}
\label{fig:back}
\end{figure}

\section{Application to Charged Particle Astronomy}
\label{sec:cpa}
Taken together, the variations of model assumptions explained in the last section lead
to an ensemble of models for the Galactic magnetic field, each of
which is compatible with the current data on the rotation measures
from extragalactic radio sources and the synchrotron emission in the
microwave band.   This ensemble of models can be considered
as a provisional estimation of the uncertainty of our knowledge of the GMF, and used to propagate the uncertainty in our knowledge of the GMF  to any kind of calculation involving the magnetic field of the Galaxy, by repeating the
calculation for each of the models.  The variations of the results give
a lower limit on the propagated uncertainty, in the absence of further input to select among or discard some of the GMF model variations.

As an example we present here the uncertainties in the arrival
direction of ultrahigh energy cosmic rays, induced by uncertainties of
the coherent component of the GMF. (See \cite{2015arXiv150804530F} for
a discussion concerning the impact random component of the GMF on
deflections and \cite{2016APh....85...54E} for a comparison of
deflections using two models of the coherent GMF.)  For this purpose
we performed a backtracking of charged particles through each of the
GMF models listed in Table~\ref{tab:cohmodels}. The results are shown
in Fig.~\ref{fig:back} for particle rigidities $R$ (rigidity =
energy/charge) of 10, 20, 30 and 60~EV.  As can be seen, for very
large rigidities (e.g.\ protons with energies of $6\times
10^{19}$~eV), the overall amount of deflection and correspondingly
also the model differences are small, confirming the long-speculated
possibility of charged particle astronomy with protons at ultrahigh
energies.  Even for rigidities as low as 20 EV, the different
deflections are mostly confined within well-defined regions so it
seems plausible that a correction for the spatially varying average
deflection based on all models, can still be used to enhance the
capabilities to identify the sources of ultrahigh energy cosmic
rays. For still lower rigidities, e.g.\ 10 EV, the differences in the
backtracked directions start to diverge considerably, but even in this
case, a more sophisticated analysis can reduce the current
uncertainties in studies of the source direction of ultrahigh energy
cosmic ray nuclei.  Further progress on constraining the magnetic
field is also expected.

\section*{Acknowledgments}
{\small We would like to thank Tess Jaffe for providing the simulations for
the cosmic-ray electrons models from~\cite{2016A&A...596A.103P}.  MU acknowledges
the financial support from the EU-funded Marie Curie Outgoing
Fellowship, Grant PIOF-GA-2013-624803 and would like to thank CCPP/NYU
for their hospitality. The research of GRF is supported in part by
the U.S.\ National Science Foundation (NSF), Grant NSF-1517319.}

\begin{multicols}{2}
\setlength\columnsep{3pt}
\bibliographystyle{na61Utphys}
\bibliography{mf}
\end{multicols}
\end{document}